\documentclass[twocolumn]{aastex631}
\usepackage{amsmath}

\shorttitle{Star formation in AGN disk and metallicity enrichment}
\shortauthors{Fan et al.}

\graphicspath{{./}{figures/}}

\begin{document}

\title{In-situ star formation in accretion disk and explanation for correlation between black hole mass and metallicity in AGNs} 

\author[0000-0001-7350-8380]{Xiao Fan}
\affiliation{Department of Astronomy, School of physics, Huazhong University of Science and Technology, Luoyu Road 1037, Wuhan, China}

\author[0000-0003-4773-4987]{Qingwen Wu$^*$}
\email{* Corresponding author: qwwu@hust.edu.cn} 
\affiliation{Department of Astronomy, School of physics, Huazhong University of Science and Technology, Luoyu Road 1037, Wuhan, China}

\begin{abstract}
Recent observations show that the metallicity of the broad line region ($Z_{\rm BLR}$) in active galactic nuclei (AGNs) is solar-to-supersolar, which is positively correlated with the mass of supermassive black holes ($M_{\rm BH}$) and does not evolve with redshift up to $z \sim 7$. We revisit the $M_{\rm BH}-Z_{\rm BLR}$ correlation with more AGNs with $M_{\rm BH}\sim 10^{6-8} M_{\odot}$ and find that the positive correlation become flat in low-mass range. It is known that outer part of accretion disks is gravitationally unstable and can fragment into stars. Considering the star formation and supernovae (SNe) in the outer AGN disk, we calculate the metal enrichment and find that positive $M_{\rm BH}-Z_{\rm BLR}$ correlation can be roughly reproduced if the stellar mass distribution is ``top-heavy". We find that the observed BLR size is more or less similar to the self-gravity radius of the AGN disk, which suggests that the BLR may be closely correlated with the underlying accretion process.

\end{abstract}


\section{Introduction} \label{sec:intro}

Most galaxies harbor a supermassive black hole (SMBH) at their nuclei, where the BH mass is around $10^{5-10} M_{\odot}$ from dwarf galaxies to massive elliptical galaxies. The active galactic nuclei (AGNs) are the most luminous persistent sources in the universe and have been detected at redshift up to $z=7.64$, which are powerful tools to study the evolution of SMBHs and galaxies \citep{feige2021,Mortlock2011,Banados2018,matsuoka2019,yang2020}. The big blue bump as observed in many luminous AGNs provides a strong evidence for mass accretion onto SMBHs \citep[e.g.,][]{shakura,Ho2008}. A large fraction of AGNs show both broad emission lines (BELs) with full width at half maximum (FWHM) of a few thousands $\rm{km\ s^{-1}}$ and narrow lines with FWHM of a few hundreds $\rm{km\ s^{-1}}$ \citep[see the reviews by][]{osterbrock1986,netzer1990,sulentic2000,czerny2019review}. The BEL is radiated from the fast moving clouds surrounding the SMBHs (e.g., broad line region, BLR), and therefore, it provides fundamental clues for the kinematics, metallicity, ionization, and structure of the nuclear regions in AGNs \citep[e.g.,][]{netzer1993,alexander1997,collin2001,wjmsfd2012,wjmblr2017,czerny2015,czerny2016,czerny2017}.

The chemical evolution of AGNs and galaxies across cosmic time contains crucial information about the star formation history and galaxy evolution \citep[see a recent review by][]{maiolino2019}. Photoionization calculations suggest that ratio of some metal emission lines or helium recombination lines (e.g., (Si{\footnotesize{IV}}+O{\footnotesize{IV}}])/C{\footnotesize{IV}}, Si{\footnotesize{III}}/C{\footnotesize{IV}}, Al{\footnotesize{III}}/C{\footnotesize{IV}}, N{\footnotesize{V}}/C{\footnotesize{IV}}, and N{\footnotesize{V}}/He{\footnotesize{II}}) can be used to infer the metallicity of BLR \citep{HF1993,hamann2002,nagao2006,tang2019,wang2022}. One of the most important results is that the BLR metallicity is generally supersolar (e.g., several $Z_{\odot}$ and up to $15Z_{\odot}$) and approximately redshift-independent \citep{HF1999,warner2004,nagao2006,jiang2007,xu2018,onoue2020,marzena2021,wang2022}, which is much higher than the metallicity of narrow line regions (NLRs) and host galaxies \citep{tremonti2004,erb2006,zahid2014,zahid2017,thomas2019,curti2020}. It is found that BLR metallicity is positively correlated with the BH mass (so-called mass-metallicity relation, or $M_{\rm BH}-Z_{\rm BLR}$ relation) or luminosity \citep{shemmer2004,warner2004,nagao2006,matsuoka2011,xu2018,lai2022}. \citet{matsuoka2011} suggested $M_{\rm BH}-Z_{\rm BLR}$ relation may be more fundamental if considering line ratios that not including nitrogen. \cite{xu2018} found that the BLR metallicity is significantly larger than the metallicity of host galaxies, which can not be explained by observational uncertainties. This is also a similar case even for a quasar at redshift of $z\sim7.5$ that the metallicity of BLR is very high but the metallicity of the host galaxy is still near solar value \citep{novak2019}. \citet{simon2010} found that the BLR metallicity does not correlate with star formation rates (SFRs) of host galaxies as constrained from far-infrared emission, which indicates that the chemical enrichment should be associated with the rapid and intense star formation in the compact nuclear region rather than transporting from the host galaxy.

The accretion disk is optically thick and geometrically thin in bright AGNs \citep[i.e., SSD,][]{shakura}. It is well-known that SSD is gravitationally unstable and prone to be self-gravitated at distances larger than several thousand Schwarzschild radii \citep{toomre1964,paczynski1978,shlosman1987,goodman2003,thompson2005}. Fragmentations due to the gravitational instability have been proposed as a mechanism for star formation in AGNs \citep[][]{cheng1999,collin1999,nayakshin2006,collin2008,wang2010,wjmsfd2011,wjmsfd2012,wada2008,wada2009}, e.g., to explain the stars in the Galactic center \citep{levin2003,nayakshin2005,levin2007,davies2020,mapelli2012}. The evolution of massive stars may lead to heavy-element enrichment near Sgr A*, which has indeed been detected in the spectra of some stars within 0.5 pc of the Galactic center \citep{do2018}. Based on SNe in the accretion disk, \citet{wang2010} found that it can roughly reproduce the observed positive metallicity–luminosity correlation. The stellar evolution in the disk environments may be much different from that in host galaxies \citep[e.g., so-called ``accretion-modified stars(AMS)" in][]{wjm2021} and potentially increase supernova rate and the number of compact objects in the nuclei of galaxies \citep{cantiello2021,dittmann2021,jermyn2021,jermyn2022}.

The BH mass-metallicity correlation may be more fundamental in AGNs \citep{warner2004,matsuoka2011}, the main goal of this work is to explore the possible metal enrichment based on the model of in-situ star formation in the outer region of the SSD. We introduce our sample to revisit the $M_{\rm BH}-Z_{\rm BLR}$ relation in Section \ref{sec:2}. In Section \ref{sec:3}, we describe the metal enrichment model. The main results are presented in section \ref{sec:4}. Discussions and conclusions are presented in Section \ref{sec:5}. Throughou this paper, we adopt flat $\Lambda$CDM cosmology with $H_0 = 70 {\rm km \ s^{-1}}$, $\Omega_{\Lambda}=0.7$ and $\Omega_{\rm m}=0.3$.

\section{Metallicity estimation and AGN Sample} \label{sec:2}

The flux ratios of high-ionization lines are commonly assumed to indicate the metallicity of the BLR in luminous AGNs. \citet{hamann2002} and \citet{nagao2006} adopted the locally optimally emitting clouds model \citep{baldwin1995} and used the photoionization code C{\footnotesize{LOUDY}} \citep{ferland2000} to simulate emission line fluxes at different metallicities. A handful of diagnostic ratios in the UV and optical wavebands have been tested to constrain the metal contents of BLRs, such as N{\footnotesize{V}}/C{\footnotesize{IV}}, (Si{\footnotesize{IV}}+O{\footnotesize{IV}})/C{\footnotesize{IV}}, Al{\footnotesize{III}}/C{\footnotesize{IV}}, N{\footnotesize{V}}/He{\footnotesize{II}}, and Fe{\footnotesize{II}}/Mg{\footnotesize{II}}. In this work, we adopt N{\footnotesize{V}}/C{\footnotesize{IV}} as the metallicity indicator, which is also wildly used in the literature \citep{HF1993,shemmer2004,shin2013,du2014,shin2019,wang2022,lai2022}, since that N{\footnotesize{V}} and C{\footnotesize{IV}} emission lines are normally prominent in the AGN spectrum. The metallicity of selected sources is estimated from the ratio of N{\footnotesize{V}}/C{\footnotesize{IV}} based on the simulations of \citet{nagao2006}.

It should be noted that there are two definitions to describe the metallicity in the literature, which are metal mass fraction ($Z$) and elemental abundance ratio by number relative to solar value ($A$). The ratio of $Z/Z_{\odot}$ is roughly the same as the value of $A$ if the abundance is around solar or sub-solar, while they differ a lot when $A>3$ (e.g., $\sim$30\% for $A=10$, see Appendix \ref{sec:Appendix A} and \citealt{hamann2002} for more details). The difference between $A$ and $Z$ comes from the fact that metal elements are produced by the fusion of hydrogen. The number of metal atoms will increase as that of hydrogen atoms decreases, where the number ratio of metal atoms to hydrogen ($A$) will not increase with $Z$ linearly. The widely adopted emission line flux ratios normally give the elemental abundance ratio rather than the metal mass fraction based on the C{\footnotesize{LOUDY}} code \citep[e.g.,][]{hamann2002,nagao2006,wang2022}. In this paper, we adopt the commonly used definitions of $Z$ and/or $A$ to denote metallicity, the correlation between them is shown in Appendix \ref{sec:Appendix A}.

The BLR metallicity has been widely explored in the literature, where the metallicity is positively correlated with BH mass when $M_{\rm{BH}} \gtrsim 10^8M_{\odot}$ \citep{warner2004,matsuoka2011,xu2018,wang2022,lai2022}. But it is still unknown whether this correlation can extend to low-mass BH sources (e.g., narrow-line Seyfert 1, NLS1). To explore the possible physical mechanism behind the $M_{\rm BH}-Z_{\rm BLR}$ correlation, we assemble a sample containing 174 AGNs from six published AGN samples, including quasars, Seyferts, and NLS1s. Among the 174 sources, 98 are low-redshift PG quasars and Seyfert 1 galaxies (${z}<0.5$) observed by International Ultraviolet Explorer or Hubble Space Telescope \citep{shin2013,du2014}. The other 41 sources with $z\sim3$ \citep{shemmer2004,shin2019} and 35 sources with ${z}\sim 6$ \citep{lai2022,wang2022} are also included in our sample. Redshift, BH mass, N{\footnotesize{V}} flux, C{\footnotesize{IV}} flux, and bolometric luminosity of these sources are all available in the literature and listed in Table \ref{table:1}. In our sample, the BH mass roughly spans five orders of magnitude from $10^6M_{\odot}$ to $10^{11}M_{\odot}$ and the redshift range from 0 to 6. It should be noted that our sample is not complete, but it will not affect our main conclusion since there is roughly no metallicity evolution with redshift in BLRs \citep{jiang2007,onoue2020,wang2022}.

\section{Model} \label{sec:3}
\begin{figure}
\centering
\includegraphics[scale=0.54]{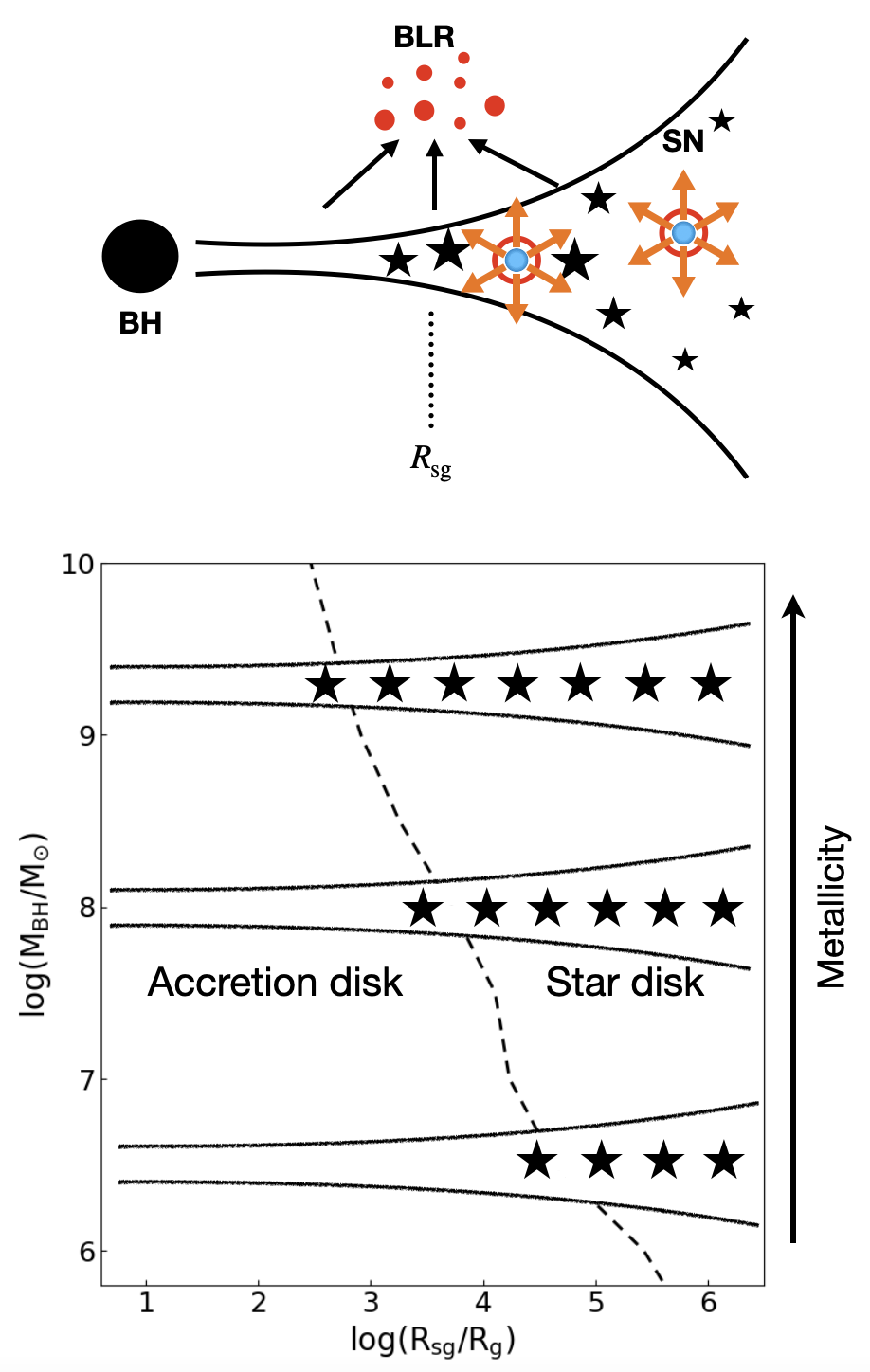}
\caption{Top panel: sketch map of our model. The inner part within $R_{\rm{sg}}$ is the standard accretion disk and the outer part is the disk with stars embedded in. Disk stars enrich the metallicity of the disk via stellar winds or SNe. The winds from accretion disk (or even some faction of gas from very energetic SN explosions can break out the disk before complete mixing) will provide metal-rich gas for BLRs. Bottom panel: sketch map for the size of $R_{\rm{sg}}$ as a function of black hole mass (dashed line) for typical accretion rate of $\dot{M}_{\rm BH}=0.1\dot{M}_{\rm Edd}$ ($\dot{M}_{\rm Edd}$ is Eddington accretion rate) and viscosity parameter of $\alpha = 0.1$. AGNs with more massive BHs have larger star disks and more stars embedded in the disk, which will lead to higher metallicity.}
\label{fig:sketchmap}
\end{figure}

The gravitational instability in the outer part of the AGN disk can trigger star formation, where is also a candidate region for producing the metals \citep[e.g.,][]{wang2010,wjmsfd2011,wjmsfd2012,marzena2021}. We explore metal enrichment by considering the contribution of stars and SNe in the outer part of the SSD. We assume that the disk with Toomre's stability parameter, $Q\equiv c_{\rm{s}}\Omega /\pi G \Sigma < 1$, is gravitationally unstable \citep{toomre1964} and the self-gravitating clouds will collapse to form stars. In Figure \ref{fig:sketchmap}, we present the sketch map of our model. As shown in the top panel, the region within the critical self-gravity radius ($R_{\rm sg}$, defined as the radius where $Q=1$) is the accretion disk while the outer part is the star disk. The stars embedded in the disk will grow through accretion, and some of them will explode as SNe. The SN ejecta mix with disk material and enrich the gas metallicity of the disk. If the SN explosions have very high kinetic energy, part of the gas will break out the disk directly \citep{moranchel2021,grishin2021,zhu2021}. In the standard disk model, heavier SMBHs predict a smaller critical self-gravity radius at the given accretion rate (e.g., the dashed line in bottom panel of Figure \ref{fig:sketchmap}). Therefore, the heavier SMBH in AGNs with more intense star formation will possibly lead to higher metallicity. We calculate the metal enrichment in more detail as follows, where the BLR gas from the wind of enriched disk and some fraction of gas pushed out by energetic SN explosions are considered.

\subsection{Star formation in accretion disk} \label{sec:3.1}
We adopt the steady-state accretion disk model proposed by \citet{sirko2003}, where the inner part of the disk ($r<R_{\rm sg}$) is the classical standard $\alpha$-disk \citep{shakura} and the outer disk ($r>R_{\rm sg}$) maintain $Q=1$ due to the possible auxiliary heating mechanisms (e.g., energy released by nuclear fusion or SNe). The disk structure can be solved for given parameters of BH mass ($M_{\rm BH}$), accretion rate ($\dot{M}_{\rm BH}$) and viscosity parameter ($\alpha$). The gas surface density of the outer disk is given by
\begin{equation}
\Sigma_{\rm{gas}} = \frac{\Omega}{\pi G} \left ( \frac{G\dot{M}_{\rm BH} }{3 \alpha} \right ) ^{\frac{1}{3}},
\end{equation}
where $G$ is gravitational constant and $\Omega=(GM_{\rm{BH}}/r^3)^{1/2}$ is Keplerian angular velocity. For the opacity of the disk, we use the opacity tables of \citet{iglesias1996} for high temperatures\footnote{\url{https://opalopacity.llnl.gov/}} and those of \citet{Ferguson05} for low temperatures\footnote{\url{https://www.wichita.edu/academics/fairmount_college_of_liberal_arts_and_sciences/physics/Research/opacity.php}} with $X=0.7$, $Z=0.02$ when disk temperature is higher and lower than $10^{3.75}\rm K$ respectively.

Following \citet{wang2010}, we estimate the star formation rate of the outer disk based on the well-known Kennicutt–Schmidt law \citep{Kennicutt1998}, 
\begin{equation}
{\dot{\Sigma}_{*}}={\rm C} \left ( \frac{\Sigma_{\rm{gas}}}{M_{\odot}\rm{pc^{-2}}} \right )^{\gamma} {M_{\odot}\rm{yr^{-1}kpc^{-2}}},
\end{equation}
where $\dot{\Sigma}_{*}$ is the surface density of star formation rate (SFR). We adopt the index $\gamma \sim 1.14$ and the constant ${\rm C}=10^{-3.41}$ that derived from the nuclear region of four nearby low-luminosity AGNs based on high-resolution observations \citep[][]{casasola2015}. The SFR of the entire star disk can be derived by $\dot{R}_{*}=\int_{R_{\rm{sg}}}^{R_{\rm{out}}} 2\pi r \dot{\Sigma}_{*}(r)dr$, where $R_{\rm{out}}$ is the outer radius of the disk. It should be noted that the size of outer radius of the star disk is unknown. In this work, we consider the BLR metal enrichment by the stars in the disk through possible disk winds and, therefore, simply adopt $R_{\rm out} \sim 10^5 R_{\rm g}$, where the viscous timescale of disk is comparable to the typical lifetime of AGNs \citep[e.g., $\sim 10^7$ years, ][]{lifetime2,lifetime1}. Furthermore, \cite{sirko2003} found that the disk cannot be much larger than $10^5R_{\rm{g}}$ to avoid the prominent excess of infrared emission.

\subsection{Metal enrichment via disk stars} \label{sec:3.2}
The star formation and evolution in the accretion disk will naturally lead to the metal enrichment due to SNe in the last stage of stellar evolution for massive stars. The chemical enrichment of a single SN is determined by the mass of the progenitor star. To estimate the contribution of the matals from the SN, we adopt the mass of the elements for each SN of zero-metallicity stars from \citet{nomoto2013}, where the remnant mass ($M_{\rm{rem}}$) and ejected mass of each element that newly produced are provided in a compiled nucleosynthesis table (Yields Table 2013\footnote{\url{http://star.herts.ac.uk/~chiaki/works/YIELD_CK13.DAT}}). The total mass of metals ejected by SNe ($M_{Z}^{\rm ej}$) can be obtained by summing the mass of the each metal element \citep[see Figure 4 of][]{nomoto2013}. Some of the metals are stored in the compact remnants of stars and will not contribute to the chemical enrichment. From this table, the mass fraction of the ejected metals to its progenitor star ($M_*$), $Z_{\rm{star}}=M_{Z}^{\rm ej}/M_*$, and mass fraction of remnant to its progenitor star, $f_{\rm{rem}}=M_{\rm rem}/M_*$, can be derived for a given mass of the progenitor star. By assuming the stellar initial mass function (IMF) characterized by a single power law function of $\phi(M_*)\propto M_*^{-\Gamma}$, the average metal mass yielded by disk stars per unit solar mass can be derived as $\Bar{Z}_{\rm{star}}= \int_{M_{\rm{min}}}^{M_{\rm{max}}} Z_{\rm{star}}(M_*)M_* \phi(M_*)dM_* $, where the IMF is normalized as $\int_{M_{\rm{min}}}^{M_{\rm{max}}} M_*\phi(M_*)dM_*=1M_{\odot}$ with minimum star mass of $M_{\rm{min}}=0.9M_{\odot}$ and maximum star mass of $M_{\rm{max}}=140M_{\odot}$ \citep[e.g.,][]{nomoto2013}. Similarly, the average mass fraction of remnants can be derived as $\Bar{f}_{\rm{rem}}=\int_{M_{\rm{min}}}^{M_{\rm{max}}} f_{\rm{rem}}(M_*)M_*\phi(M_*)dM_*$.

\citet{wjmsfd2011} suggested that the mixing timescale of metals with their surroundings is shorter than hundreds of years, which means SN ejecta can mix with disk gas sufficiently. The metallicity of the disk at a given time can be described as $Z(t) = M_{Z}(t)/M_{\rm gas}$, where $M_{Z}(t)$ is the total mass of metals by the time $t$ and $M_{\rm{gas}}=\int 2\pi r{\Sigma}_{\rm{gas}}(r)dr$ is the total gas mass of the disk. Following \cite{wang2010}, we calculate the net increase of the total mass of metals in the disk as follows,
\begin{equation} \label{equation:enrich}
\begin{split}
Z(t)M_{\rm{gas}}=\int_{0}^{t}(&Z_{0}\dot{M}_{\rm{out}}+\Bar{Z}_{\rm star}\dot{R}_{*}\\
&-Z(\tau)\dot{M}_{\rm{BH}}-Z(\tau)\Bar{f}_{\rm{rem}} \dot{R}_{*})d\tau,
\end{split}
\end{equation}
where the first term on the right side of the equation is the initial metal of supplied gas from the outer boundary of the accretion disk ($Z_{0}$ is the metal mass fraction of the initial gas and $\dot{M}_{\rm out}=\dot{M}_{\rm BH}+\Bar{f}_{\rm rem} \dot{R}_{*}$ is the accretion rate at the outer boundary), the second term represents the metals contributed by the SNe, the third and the last term represent the consumption of the metals that advected into SMBHs and stored in compact remnants respectively.

For a steady star disk, the total mass of the disk roughly remains constant while the metal mass increases, which causes the metallicity of the disk increase with time. The solution of equation \ref{equation:enrich} is given by \citep[][]{wang2010}
\begin{equation} \label{equ:5}
    Z(t)=Z_{\rm{max}}+(Z_{0}-Z_{\rm{max}})\rm{exp}(-{\it t}/{\it t_{\rm z}}),
\end{equation}
where $t_{\rm z}=M_{\rm{gas}}/(\dot{M}_{\rm{BH}}+\Bar{f}_{\rm{rem}}\dot{R}_{*})$. It can be seen that when $t>t_z$, the metallicity of the disk will approach a saturation value, 
\begin{equation} \label{equa:6}
Z_{\rm{max}}=\frac{Z_{0}\dot{M}_{\rm{out}}+\Bar{Z}_{\rm star}\dot{R}_{*}}{\dot{M}_{\rm{BH}}+\Bar{f}_{\rm{rem}} \dot{R}_{*}}.
\end{equation}

\begin{figure}
\centering
\includegraphics[scale=0.42]{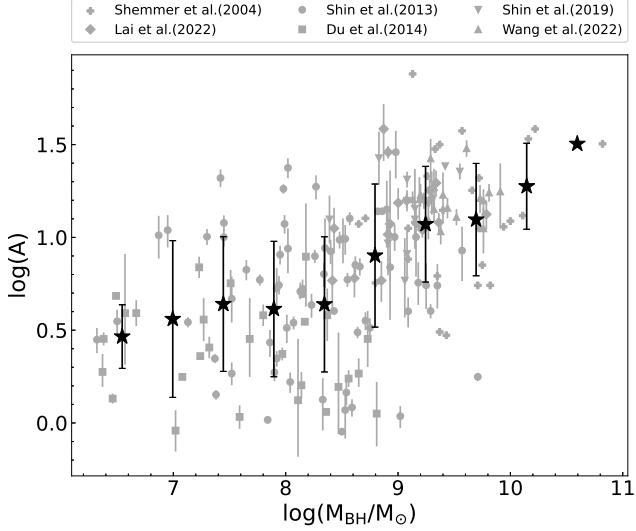}
\caption{Elemental abundance ratio versus SMBH mass. The grey points with different symbols indicate sources assembled from different literature \citep{shemmer2004,shin2013,du2014,shin2019,lai2022,wang2022}. The stars indicate the average value of each mass bin and the error bars represent the standard deviation.}
\label{fig:observation}
\end{figure}

\begin{figure}
\centering
\includegraphics[scale=0.58]{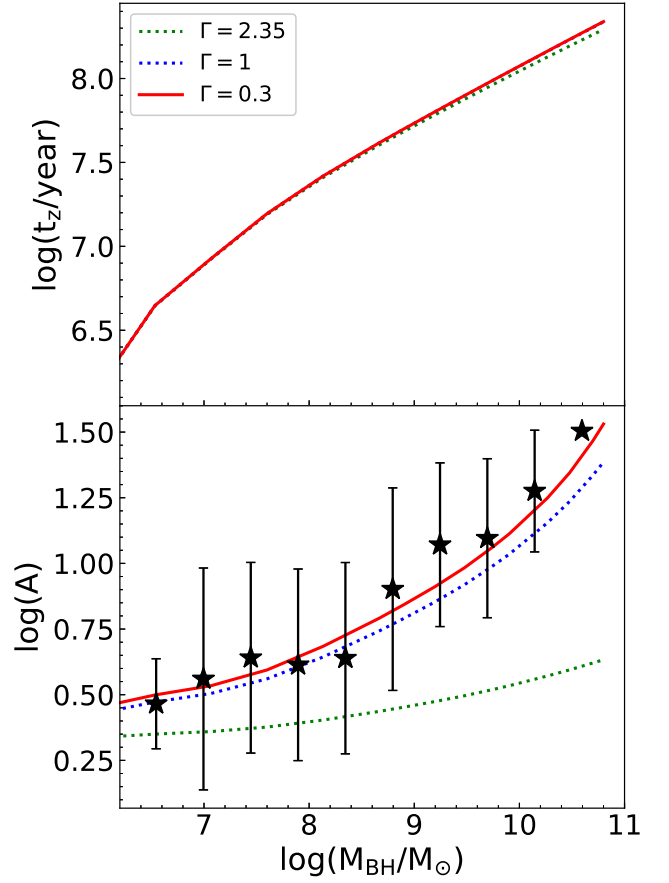}
\caption{Top panel: the typical time $t_{\rm{Z}}$ for the metallicity of the disk to approach the saturation value (Equation \ref{equ:5}). Bottom panel: model prediction for the abundance ratio and comparison with the average observational results, where different lines correspond to different stellar mass distribution indices as shown in the legend.}
\label{fig:result_MZ}
\end{figure}

Part of the gas will directly escape from the accretion disk before completely mixture with disk materials if the SN explosion is energetic enough or close to the skin of the disk. Based on simulations of a mid-plane SN explosion, the ejecta mass may include both disk gas ($M_{\rm{BLR,disk}}$) and SN ejecta ($M_{\rm{BLR,SN}}$), where the ratio $\epsilon=M_{\rm{BLR,disk}}/M_{\rm{BLR,SN}}\sim 10$ for a typical mid-plane SN explosion with ejecta mass $M_{\rm{ej}}=10M_{\odot}$ and $E_{\rm{SN}}=2 \times 10^{52} \ \rm{erg}$ \citep[][]{moranchel2021}. In this case, the metallicity of BLR can be written as
\begin{equation} \label{equation:ZBLR}
    Z_{\rm{BLR}}=\frac{\Bar{Z}_{\rm{SN}}+\epsilon Z_{\rm{max}}}{1+\epsilon},
\end{equation}
where $\Bar{Z}_{\rm{SN}}= \int_{M_{\rm{crit}}}^{M_{\rm{max}}} Z_{\rm{SN}}(M)M\phi(M)dM$ is metal mass fraction of SN ejecta averaged by stellar mass distribution and $M_{\rm{crit}}=11M_{\odot}$ is the critical mass for stars to produce SNe \citep[][]{nomoto2013}. 

\section{Results}\label{sec:4}

\subsection{$M_{\rm BH}-Z_{\rm BLR}$ correlation and physical explanation}\label{sec:4.1}

We estimate the elemental abundance ratio from the line ratio of N{\footnotesize{V}}/C{\footnotesize{IV}} for our sample as described in Section \ref{sec:2}, and results are presented in Figure \ref{fig:observation}. Based on our sample, the elemental abundance ratio and BH mass are positively correlated with Spearman correlation coefficient of $r_{\rm{s}}=0.58$ ($p=2.30 \times 10^{-19}$). Due to the large scatter, we present the average abundance ratios at given 10 mass bins (stars with error bars in Figure \ref{fig:observation}). It can be found more clearly that the abundance is positively correlated with the BH mass for the sources with $M_{\rm BH} \gtrsim 10^8M_{\odot}$, and this correlation become flatter for the sources with $M_{\rm BH} \lesssim 10^8M_{\odot}$. The Spearman correlation coefficient is $r_{\rm s}=0.51$ ($p=2.59 \times 10^{-10}$) for the sources with $M_{\rm BH} > 10^8M_{\odot}$ and $r_{\rm{s}}=0.16$ ($p=0.31$) for $M_{\rm BH} < 10^8M_{\odot}$ respectively.

\begin{figure}
\centering
\includegraphics[scale=0.38]{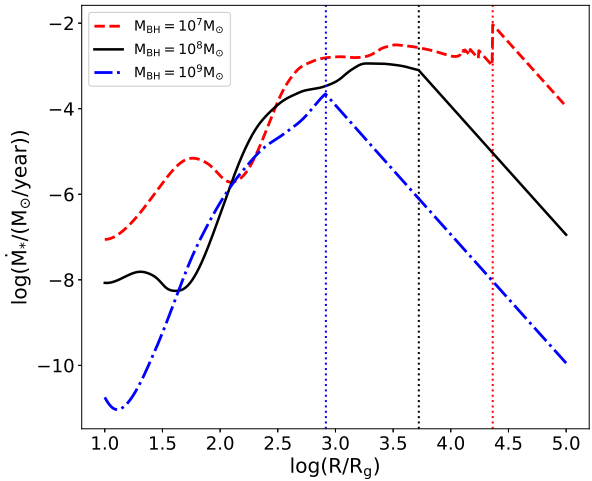}
\caption{The illustration of the accretion rate of disk stars with mass $M_{*}=1M_{\odot}$ at different radii. The solid, dashed and dash-dotted lines represent different BH masses as shown in the legend. Dotted vertical lines represent location of $R_{\rm{sg}}$ by assuming $\dot{M}_{\rm BH}=0.1\dot{M}_{\rm Edd}$ and $\alpha=0.1$.}
\label{fig:star_accretion}
\end{figure}

In the top panel of Figure \ref{fig:result_MZ}, we present the typical timescale $t_{\rm Z}$ for the elemental abundance ratio of the disk to approach the saturation value (see equation \ref{equ:5}). The timescale is normally less than 0.1 Gyr even for $M_{\rm BH}=10^{10}M_{\odot}$, while the typical value is $\sim 20$ Myr for $M_{\rm BH}=10^{8}M_{\odot}$. The short saturation timescale may account for redshift-independent of BLR metallicities. By assuming the stellar mass distribution of stars embedded in the disk, the elemental abundance ratio of the BLR gas can be derived (equation \ref{equation:ZBLR} and \ref{A5}). In the bottom panel of Figure \ref{fig:result_MZ}, we present a comparison of the theoretical model with the average observational results. The solid, dashed and dotted lines represent results with the stellar mass distribution indices of $\Gamma = 0.3$, $\Gamma=1 $ and $\Gamma = 2.35$ (the Salpeter IMF) respectively, where $\alpha=0.1$, the typical accretion rate of $\dot{M}_{\rm BH}=0.1\dot{M}_{\rm Edd}$ for our samples, and initial metallicity of $Z_0=Z_{\odot}$ \citep[e.g., the typical metallicity of NLRs,][]{castro2017} are adopted. It can be found that the prediction with Salpeter IMF is much flatter than that of observational results. A ``top-heavy" stellar mass distribution (e.g., $\Gamma<1$) is required to reproduce the average observational results. It should be noted if SNe is close to the skin of the disk or the inner region of disk with lighter SMBHs (e.g., $<10^7M_{\odot}$), the SN ejecta can break out from the disk directly without fully mix with the disk gas (i.e., the low $\epsilon$, \citealt{grishin2021,moranchel2021}), which can lead to a quite high BLR metallicity (e.g., $A\sim30$, comparable to that of SN ejecta).

\subsection{The possible SNe region and BLR size}\label{sec:4.2}

The BLR gas is most possibly related to the accretion disk, star disk, and/or torus even though its origin remains unknown. The stars embedded in the disk can grow up quickly by accreting the surrounding gas (e.g., it takes $\sim 10^{4-6}$ years for stars at $R_{\rm sg}$ grow from $1 M_{\odot}$ to $100 M_{\odot}$), which is a possible mechanism for the ``top-heavy" stellar mass distribution \citep{toyouchi2022}. We estimate the accretion rate of disk stars located at different radii of AGN disk, where the Bondi-like accretion is adopted
\begin{equation}
    \dot{M}_{*}=\eta \pi R^{2}_{\rm{*B}} \rho c_{\rm s},
\end{equation}
where $R_{\rm{*B}}=2GM_{*}/c_{\rm{s}}^2$ is Bondi radius of embedded stars with mass $M_{*}$, $\rho$ is mass density and $c_{\rm{s}}$ is the sound speed of the surrounding material in AGN disk. It should be noticed that the accretion rate may be affected by radiative feedback, actual accretion radius (e.g., $R_{\rm B}$ may be less than disk thickness), tidal effect, etc \citep{dittmann2021}. Following \citet{Derdzinski2022}, we adopt the empirical efficiency factor $\eta=0.01$, which will not affect our main conclusions. In Figure \ref{fig:star_accretion}, we present the accretion rate for a star with mass of $M_{*}=1M_{\odot}$ at different radii for three typical masses of SMBHs in our disk model. Disk stars located at different radii accrete at different rates, which depend on the local disk environments. It can be found that the accretion rate of stars peaks at about several thousand to ten thousand gravitational radii, where the accretion rate increases as decreases of radius, which will approach the maximal value at $r\sim R_{\rm sg}$. The mass of the star can grow to several hundred solar masses within million years if the star is embedded in a region at $r\sim R_{\rm sg}$, which will lead to SNe. Considering the star migration, the probability for the SN of the disk star may be highest in the region around $R_{\rm sg}$.

\begin{figure}
\centering
\includegraphics[scale=0.4]{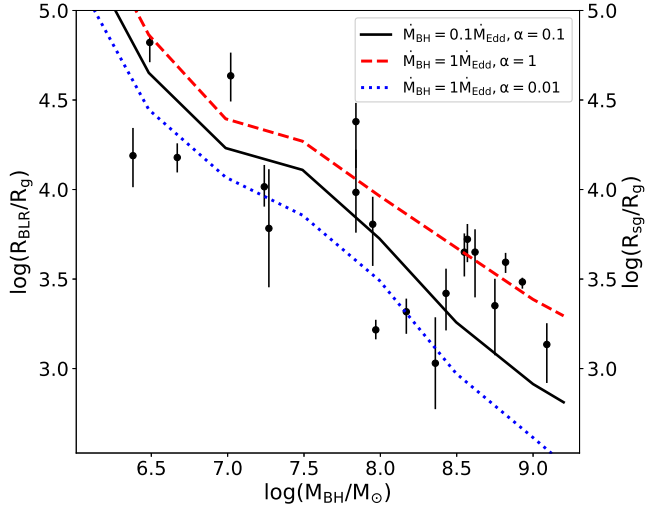}
\caption{The relation between BLR size and BH mass for 20 sources with directly reverbration mapping measurements. The dot points with error bars are the observation of BLR size. The solid, dashed and dotted lines represent the size of $R_{\rm{sg}}$ for given BH masses with typical value of accretion rate $\dot{M}_{\rm BH}=0.1\dot{M}_{\rm Edd}$, viscosity $\alpha=0.1$; $\dot{M}_{\rm BH}=1\dot{M}_{\rm Edd}$, $\alpha=1$; and $\dot{M}_{\rm BH}=0.01\dot{M}_{\rm Edd}$, $\alpha=0.01$ respectively.}
\label{fig:Rsg_Rblr}
\end{figure}

Among our sample, 20 sources have time-delay measurements ($\tau_{\rm{obs}}$) from reverberation mapping observations (see table \ref{table:1}), which provide the measurementss of the BLR size ($R_{\rm{BLR}}=c \tau_{\rm{obs}}$) directly. In Figure \ref{fig:Rsg_Rblr}, we present the relation between the BH mass and BLR size for these 20 sources. For comparison, we also present the theoretical size of $R_{\rm sg}$, where different viscosity parameters and accretion rates are considered. The observational anti-correlation is roughly consistent with the size of $R_{\rm sg}$ for typical value of viscosity $\alpha\sim0.01-1$ and $\dot{M}_{\rm BH}/\dot{M}_{\rm Edd}=0.01-1$. 

\section{Conclusions and Discussions} \label{sec:5}

The BLR metallicity of AGNs is generally super-solar, positively correlates with $M_{\rm BH}$, and is roughly independent of the redshift. These observations suggested that the metal elements are most possibly produced in the nuclear region close to SMBHs. The stars that formed in the outer region of SSD or are captured by accretion disks will evolve quickly in these gas-rich environments and disk stars will provide considerable metal elements through SNe. In this work, we revisit the possible correlation between the metallicity of BLR and BH mass with more AGNs with $M_{\rm BH}\sim 10^{6-8} M_{\odot}$ based on N{\footnotesize{V}}/C{\footnotesize{IV}} as a metallicity indicator and find that positive $M_{\rm BH}-Z_{\rm BLR}$ correlation become flat at low-mass range. Based on the assumption of star formation in the self-gravitating disk, we calculate the metal enrichment and find that the positive $M_{\rm BH}-Z_{\rm BLR}$ correlation can be reproduced if the stellar mass distribution of disk stars is ``top-heavy". We find that the size of BLR is roughly consistent with the self-gravity radius, where the embedded stars grow most rapidly and the probability of producing SNe is highest. 

The positive $M_{\rm BH}-Z_{\rm BLR}$ correlation has been reported in several former works \citep{shemmer2004,xu2018,wang2022,lai2022}, where the $Z_{\rm BLR}$ depends much more significantly on $M_{\rm BH}$ rather than $L/L_{\rm Edd}$ \citep[see][for more details]{warner2004,matsuoka2011}. Some works also proposed that this correlation may be not strong \citep{shin2013,du2014}. In this work, we find that the different conclusions are mainly caused by the sources in different BH mass range, where this relation is indeed weak in low mass regime ($M_{\rm BH}\sim 10^{6-8} M_{\odot}$) but show a positive correlation when $M_{\rm BH}\gtrsim 10^{8} M_{\odot}$. The estimation of metallicity for more sources with $M_{\rm BH}\sim 10^{6-7} M_{\odot}$ and $M_{\rm BH}\sim 10^{9-10} M_{\odot}$ are wished to be further explored.

In this work, the BLR metallicity is estimated from the line ratio of N{\footnotesize{V}}/C{\footnotesize{IV}}, while the ratio of N{\footnotesize{V}}/He{\footnotesize{II}}, (Si{\footnotesize{IV}}+O{\footnotesize{IV}}])/C{\footnotesize{IV}}, and Al{\footnotesize{III}}/C{\footnotesize{IV}} are also widely adopted in the literature \citep[e.g.,][]{hamann2002,nagao2006,matsuoka2011}. \citet{matsuoka2011} found that there is a significant correlation between $M_{\rm BH}$ and $Z_{\rm{BLR}}$ as inferred from all above four metallicity-sensitive emission-line flux ratios. It should be noted that the line flux ratios of (Si{\footnotesize{IV}}+O{\footnotesize{IV}})/C{\footnotesize{IV}}, and Al{\footnotesize{III}}/C{\footnotesize{IV}} are better served as a metallicity indicator since that they are not correlated with the Eddington ratios. However, the lines of N{\footnotesize{V}} and C{\footnotesize{IV}} are normally stronger than other lines, and we adopt these two lines to estimate the BLR metallicity, even though it may be affected by the possible ionization parameter or the shape of the ionizing continuum \citep{hamann2002,nagao2006}. The metallicity estimation based on different line ratios will not affect our main results since that they are all calibrated.

The physical origin of $M_{\rm BH}-Z_{\rm BLR}$ correlation and no redshift evolution of elemental abundance ratios in AGNs are unclear. The higher metallicity in BLRs than that in NLRs and host galaxies suggest that the metal enrichment is most likely to happen in the compact nuclear region (e.g., pc or even smaller scales). The high densities and large amounts of gas in the central region of AGN likely trigger rapid star formation and enrich the metallicity of environments quickly due to SNe. We explore the possible metal enrichment process based on the model of in-situ star formation in gravitationally unstable regions in the outer accretion disk. More massive SMBHs have larger star formation regions, and more disk stars will lead to higher metallicities (see the cartoon picture of Figure \ref{fig:sketchmap}). The sound speed of the disk is lower for AGNs with smaller SMBHs, where the SN explosion can break out of the disk much easier than that of more massive BHs \citep{moranchel2021,grishin2021}. The SNe will contribute metals to the BLR directly, which will lead to a shallower $M_{\rm BH}-Z_{\rm BLR}$ correlation. Based on the well-known Kennicutt–Schmidt law \citep{Kennicutt1998}, we calculate the SFR in the outer accretion disk and estimate the metal enrichment through SNe by assuming stellar mass distribution. We find that the standard IMF with $\Gamma=2.35$ predicts a much shallower $M_{\rm BH}-Z_{\rm BLR}$ correlation, which can not explain the observed correlation (e.g., the green dashed line in the bottom panel of Figure \ref{fig:result_MZ}). Our results imply a ``top-heavy" stellar mass distribution, i.e., $\Gamma < 1$ is needed to reproduce the average $M_{\rm BH}-Z_{\rm BLR}$ relation, which is roughly consistent with the results of \citet{toyouchi2022,Qiyanqing2022}. The direct observational constraints on the stellar mass distribution in the galaxy center are difficult due to the limited telescope resolutions. The high-resolution observations of the stars near Sgr A* indicate an extremely ``top-heavy" stellar mass distribution in the galactic center (e.g., $\sim 1{''}-10^{''}$, \citealt{bartko2010}), where hundreds of massive, young OB stars and Wolf-Rayet stars have been detected in the central nuclear region (e.g., within $\sim$0.5 pc, \citealt{krabbe1995,genzel2003,ghez2003}. Based on the fragment masses of pre-stellar clumps caused by gravitational instability in the AGN disc, \citet{Derdzinski2022} also suggested that the stellar mass distribution in the nuclear region of AGNs should be ``top-heavy" with $\Gamma \sim 0.7$, which is roughly consistent with our results. The typical timescale for metallicity to approach the saturation value is 0.1 Gyr, which can roughly explain the non-redshift evolution up to $z \sim 7$.

Some metallicity-sensitive emission-line flux ratios strongly support a tight correlation between $Z_{\rm{BLR}}$ and $M_{\rm BH}$, while $Z_{\rm{BLR}}$ does not correlate with the Eddington ratios and redshift \citep[e.g.,][]{warner2004,matsuoka2011}. It should be also noted that the line ratios related to N{\footnotesize{V}} and Fe{\footnotesize{II}} are possibly correlated with the Eddington ratio. \cite{matsuoka2011} proposed that the possible correlation between N{\footnotesize{V}}/C{\footnotesize{IV}} and Eddington ratios may be caused by a delay of the black hole accretion rate relative to the onset of nuclear star formation of about $10^8$ years, which is the timescale required for the nitrogen enrichment. The Fe{\footnotesize{II}}/Mg{\footnotesize{II}} line flux ratio, the first-order proxy of Fe/$\alpha$ elemental abundance ratio, is another indicator of AGN metallicity \citep{shin2019,sameshima2017,shin2021}, which acts like a clock of star formation history since that the iron element is mainly ejected by the type Ia SN that is delayed by $10^8$ years relative to $\alpha$ elements from core-collapse SNe (Type II and Type Ib/Ic SNe). The star formation and stellar evolution in the nuclear region close to SMBH are still highly uncertain \citep{cantiello2021,dittmann2021,jermyn2021,jermyn2022,liyaping2022}. The simultaneous measurements of different line ratios can provide a chance to constrain the possible stellar evolution for the stars embedded in accretion disk. We estimate the star formation rate from the KS law with an observational index of 1.14 as constrained from nuclear regions of AGNs ($\sim$ 100 pc, \citealt{casasola2015}), and high-resolution constraints are wished to further explore the stellar properties in these special environments.

It should be noted that the origin of materials in the BLR clouds is still unclear, which may originate from the winds/envelopes of bloated stars \citep{alexander1997}, inflow from large-scale gas (e.g., torus clumps), the gas of SNe from the stars embedded in accretion disk and/or winds of the accretion disk \citep[see][]{wjmsfd2012,wjmblr2017,galianni2013,czerny2016,czerny2017,czerny2019review}. Theoretical studies have suggested that stars embedded in the self-gravitating regimes of AGN disks will suffer strong evolution after their formation. The SNe from massive stars can support a geometrically thick, turbulent gas disk where the turbulent viscosity excited by SNe transports the angular momentum of the metal-rich gas efficiently \citep[e.g.,][]{wada2002,wada2008}. The ejected material from the SNe that happened in the outer gravitationally unstable disk also enrich the metallicity of the AGN disk \citep[e.g.,][]{wjmsfd2012}, which may be the physical reason for the high metallicity of BLR if considering the BLR clouds are related to the underlying AGN disk.
The evolution of AGN stars should be much different from the classical star evolution with quite low density surroundings (or low accretion rate, e.g., \citealt{cantiello2021,dittmann2021,wjm2021,jermyn2022}). With the assumption of Bondi-like accretion, we calculate the accretion rate of the stars at different disk radii and find that the accretion rates always peak at $\sim R_{\rm{sg}}$ (see Figure \ref{fig:star_accretion}). We proposed that the stars should most possibly undergo a runaway accretion and explode as SNe in the region from $\sim 10 R_{\rm sg}$ to $\sim 0.1 R_{\rm sg}$, where the stars can increase up to several tens to several hundred solar masses by assuming the typical migration timescale of $\sim 10^{6}$ yr for type I migration \citep{Derdzinski2022}. Therefore, the metallicity in this region should be highest in the accretion disk. The BLR clouds formed from the disk winds and/or SN-ejected materials will have quite high metallicity. In Figure \ref{fig:Rsg_Rblr}, we compare the observed BLR size with the size of $R_{\rm{sg}}$ as predicted from the disk model, which is roughly consistent with each other. It should be noted that above correlation itself cannot give the physical origin of BLR, which just suggests that the BLR clouds may be closely correlated with the underlying disk. More detailed investigations considering the stellar evolution, accretion, and migration are needed to further test this issue, which will be our future work.

\

We thank our referee for constructive comments and suggestions that help us to improve this paper. We appreciate Jianmin Wang, Bozena Czerny, Luis C. Ho and the members in our BH group for very useful discussions and suggestions. This work is supported by the NSFC (grants U1931203,12233007), the science research grants from the China Manned Space Project (No. CMS-CSST-2021-A06). The authors acknowledge Beijng PARATERA Tech CO.,Ltd. for providing HPC resources that have contributed to the results reported within this paper.

\appendix

\section{elemental abundance AND METAL MASS FRACTION} \label{sec:Appendix A}

There are mainly two expressions to describe metallicities, which are customarily defined as the value compared to the Sun. The first one is the elemental abundance ratio, which is defined as 
\begin{equation} \label{A1}
A\equiv \frac{\rm{metal}/H}{(\rm{metal}/H)_{\odot}},
\end{equation}
where ``metal" and H represent the number of metal elements (heavier than helium) and hydrogen atoms per unit of volume respectively. 

\begin{figure}
\centering
\includegraphics[scale=0.45]{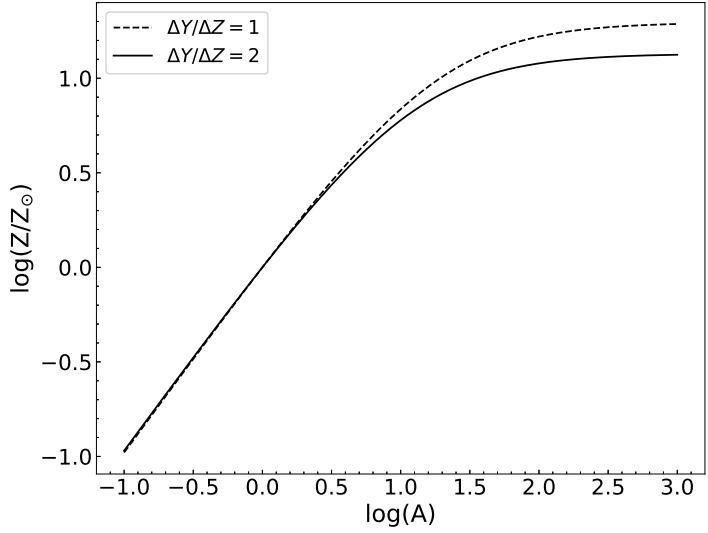}
\caption{The relation between A and Z. The solid line represents $\Delta Y/\Delta Z =2 $, which we use in this paper. The dashed line represents $\Delta Y/\Delta Z =1 $. }
\label{fig:Z-A}
\end{figure}

The second expression is defined as
\begin{equation}\label{A2}
\frac{Z}{Z_{\odot}} \equiv \frac{M_{\rm{metal}}/M}{(M_{\rm{metal}}/M)_{\odot}},
\end{equation}
where $Z$ is the metal mass fraction, $M_{\rm metal}$ is metal mass and $M$ is the total mass. The mass fractions of hydrogen and helium are defined as $X\equiv M_{\rm H}/M$ and $Y\equiv M_{\rm {He}}/M$ respectively, where
\begin{equation}\label{A3}
X+Y+Z\equiv1.
\end{equation}
Here, we assume the average metal atomic mass is $\mu$ and the number of hydrogen, helium and metal atoms are $a$, $b$ and $c$ respectively. With the relation \ref{A1}, \ref{A3}, and $\Delta Y/\Delta Z = \eta$, we have
\begin{equation}
\begin{split}
AZ_{\odot}/(uX_{\odot})=&c/a  \\
a+4b+\mu c=&1\\
\eta(\mu c-Z_{\odot})=4b&-Y_{\odot}, 
\end{split}
\end{equation}
where $X_{\odot}, Y_{\odot}, Z_{\odot} $ are the hydrogen, helium and metal mass fraction of the Sun. The relation between $A$ and $Z$ is given by
\begin{equation} \label{A5}
{\frac{Z}{Z_{\odot}}=\frac{A(1+{\eta}Z_{\odot}-Y_{\odot})}{X_{\odot}+({\eta}+1)AZ_{\odot}}}. 
\end{equation}

As shown in Figure \ref{fig:Z-A}, we present the relation between $Z/Z_{\odot}$ and $A$, where they do not follow a linear relation when $A \gtrsim 3$ and $Z/Z_{\odot}$ will saturate when $A> 30$. In this paper, we adopt typical value of $\Delta Y/\Delta Z =2 $ \citep{izotov2007} and $X_{\odot}=0.711$, $Y_{\odot}=0.27$, $Z_{\odot}=0.019$ \citep{GN1993,Asplund2009}.

\bibliography{metallicity}{}
\bibliographystyle{aasjournal}

\startlongtable
\begin{deluxetable*}{lcccccc}
\tablecaption{AGN properties and flux ratios of our sample\label{table:1}}
\tablewidth{0pt}
\tabcolsep=0.4cm
\tablenum{1}
\tablehead{
\colhead{Object} & \colhead{z} & \colhead{log$M_{\rm{BH}}$} & \colhead{log$L_{\rm bol}$} & \colhead{log(N{\footnotesize{V}}/C{\footnotesize{IV}})} & \colhead{$\tau_{\rm obs}$} & \colhead{Ref.} \\
\nocolhead{Common} & \nocolhead{Common} & \colhead{($M_{\odot}$)} & \colhead{($\rm{erg \ s^{-1}}$)} & \nocolhead{} & \colhead{(days)} & \nocolhead{}
}
\startdata
2E 1615+0611 & 0.038 & 7.59 & 44.22 & $-1.00\pm 0.07$ & \nodata  & 1 \\
Fairall 9 & 0.047 & 8.17 & 45.12 & $-0.49\pm 0.01$ & $17.4^{+3.2}_{-4.3}$  & 1, 7 \\
HB89 1028+313 & 0.178 & 8.73 & 45.31 & $-0.57\pm 0.13$ & \nodata  & 1  \\
Mrk 42 & 0.025 & 6.37 & 43.90 & $-0.73\pm 0.09$ & \nodata & 1  \\
Mrk 106 & 0.123 & 8.47 & 45.38 & $-0.82\pm 0.28$ & \nodata  & 1  \\
Mrk 110 & 0.035 & 7.02 & 44.02 & $-1.08\pm 0.12$ & $25.6^{+8.9}_{-7.2}$  & 1, 7  \\
Mrk 142 & 0.045 & 7.27 & 44.54 & $-0.48\pm 0.10$ & $6.4^{+7.3}_{-3.4}$  & 1, 7  \\
Mrk 290 & 0.030 & 7.97 & 44.53 & $-0.64\pm 0.03$ & $8.7^{+1.2}_{-1.0}$  & 1, 7 \\
Mrk 359 & 0.017 & 6.46 & 44.18 & $-0.89\pm 0.03$ & \nodata & 1 \\
Mrk 493 & 0.031 & 6.49 & 44.24 & $-0.37\pm 0.01$ & $11.6^{+1.2}_{-2.6}$ & 1, 7 \\
Mrk 618 & 0.035 & 7.51 & 44.53 & $-0.32\pm 0.05$ & \nodata  & 1 \\
Mrk 771 & 0.063 & 8.15 & 44.82 & $-0.36\pm 0.02$ & \nodata  & 1 \\
Mrk 841 & 0.036 & 8.11 & 44.75 & $-0.90\pm 0.33$ & \nodata  & 1 \\
Mrk 876 & 0.129 & 8.65 & 45.83 & $-0.74\pm 0.08$ & \nodata  & 1 \\
Mrk 1018 & 0.042 & 8.18 & 44.76 & $-0.22\pm 0.20$ & \nodata  & 1 \\
Mrk 1126 & 0.011 & 7.68 & 43.75 & $-0.57\pm 0.19$ & \nodata  & 1 \\
Mrk 1239 & 0.020 & 6.57 & 44.26 & $-0.45\pm 0.24$ & \nodata  & 1 \\
Mrk 1243 & 0.035 & 7.23 & 44.32 & $-0.26\pm 0.04$ & \nodata  & 1 \\
Mrk 1514 & 0.016 & 7.08 & 44.36 & $-0.76\pm 0.01$ & \nodata  & 1 \\
NGC 3783 & 0.010 & 7.24 & 44.10 & $-0.65\pm 0.01$ & $10.2^{+3.3}_{-2.3}$  & 1, 7 \\
NGC 4051 & 0.002 & 6.38 & 42.95 & $-0.57\pm 0.03$ & $2.1^{+0.9}_{-0.7}$  & 1, 7 \\
NGC 4593 & 0.009 & 6.67 & 43.78 & $-0.45\pm 0.06$ & $4.0^{+0.8}_{-0.7}$  & 1, 7 \\
NGC 5548 & 0.017 & 8.36 & 44.28 & $-0.97\pm 0.02$ & $13.9^{+11.2}_{-6.2}$  & 1, 8 \\
NGC 5940 & 0.034 & 7.80 & 44.31 & $-0.46\pm 0.05$ & \nodata  & 1 \\
PG 0906+484 & 0.117 & 8.37 & 44.99 & $-0.46\pm 0.12$ & \nodata  & 1 \\
PG 0923+129 & 0.029 & 7.32 & 44.34 & $-0.61\pm 0.05$ & \nodata  & 1 \\
SBS 1150+497 & 0.334 & 8.56 & 45.55 & $-0.77\pm 0.05$ & \nodata  & 1 \\
Ton 256 & 0.131 & 8.14 & 45.38 & $-0.81\pm 0.08$ & \nodata  & 1 \\
US 1329 & 0.254 & 8.81 & 46.10 & $-0.98\pm 0.19$ & \nodata & 1 \\
PG 0003+158& 0.450 & 9.25& 46.64& $ -0.33 \pm 0.08 $ & \nodata & 2 \\
PG 0003+199& 0.026 & 7.13& 44.43& $ -0.49 \pm 0.02 $ & \nodata & 2 \\
PG 0007+106& 0.089 & 8.71& 44.96& $ -0.49 \pm 0.04 $ & \nodata & 2 \\ 
PG 0026+129& 0.142 & 8.57& 45.57& $ -0.08 \pm 0.02 $ & $111.0^{+24.1}_{-28.3}$ & 2, 7 \\
PG 0049+171& 0.064 & 8.33& 44.36& $ -0.90 \pm 0.13 $ & \nodata & 2 \\
PG 0050+124& 0.061 & 7.42& 44.69& $ 0.08 \pm 0.03 $ & \nodata & 2 \\
PG 0052+251& 0.155 & 8.55& 45.78& $ -0.31 \pm 0.03 $ & $89.8^{+24.5}_{-24.1}$ & 2, 7 \\
PG 0157+001& 0.164 & 8.15& 45.70& $ -0.34 \pm 0.04 $ & \nodata & 2 \\
PG 0804+761& 0.100 & 8.82& 45.99& $ -0.05 \pm 0.01 $ & $146.9^{+18.8}_{-18.9}$ & 2, 7 \\
PG 0838+770& 0.131 & 8.13& 45.24& $ -0.35 \pm 0.03 $ & \nodata & 2 \\
PG 0844+349& 0.064 & 7.95& 45.00& $ -0.21 \pm 0.03 $ & $32.3^{+13.7}_{-13.4}$ & 2, 7 \\
PG 0921+525& 0.035 & 7.38& 44.30& $ -0.87 \pm 0.03 $ & \nodata & 2 \\
PG 0947+396& 0.206 & 8.66& 45.84& $ -0.26 \pm 0.03 $ & \nodata & 2 \\
PG 1011$-$040& 0.058 & 7.30& 44.83& $ -0.15 \pm 0.03 $ & \nodata & 2 \\
PG 1012+008& 0.185 & 8.23& 45.41& $ -0.41 \pm 0.05 $ & \nodata & 2 \\
PG 1022+519& 0.045 & 6.32& 44.48& $ -0.57 \pm 0.05 $ & \nodata & 2 \\
PG 1048+342& 0.167 & 8.35& 44.74& $ -0.19 \pm 0.11 $ & \nodata & 2 \\
PG 1049$-$005& 0.357 & 9.16& 46.17& $ -0.15 \pm 0.11 $ & \nodata & 2 \\
PG 1103$-$006& 0.425 & 9.30& 46.11& $ 0.00 \pm 0.05 $ & \nodata & 2 \\
PG 1115+407& 0.154 & 7.65& 45.62& $ -0.27 \pm 0.04 $ & \nodata & 2 \\
PG 1116+215& 0.177 & 8.51& 46.30& $ -0.15 \pm 0.08 $ & \nodata & 2 \\
PG 1119+120& 0.049 & 7.45& 44.62& $ -0.15 \pm 0.03 $ & \nodata & 2 \\
PG 1121+422& 0.234 & 8.01& 45.94& $ -0.52 \pm 0.06 $ & \nodata & 2 \\
PG 1149$-$110& 0.049 & 7.90& 44.25& $ -0.73 \pm 0.05 $ & \nodata & 2 \\
PG 1151+117& 0.176 & 8.53& 45.65& $ -0.15 \pm 0.05 $ & \nodata & 2 \\
PG 1202+281& 0.165 & 8.59& 44.95& $ -0.94 \pm 0.05 $ & \nodata & 2 \\
PG 1211+143& 0.085 & 7.94& 45.63& $ -0.33 \pm 0.03 $ & $93.8^{+25.6}_{-42.1}$ & 2, 7 \\
PG 1216+069& 0.334 & 9.18& 46.52& $ -0.32 \pm 0.08 $ & \nodata & 2 \\
PG 1226+023& 0.158 & 8.93& 46.59& $ -0.26 \pm 0.05 $ & $146.8^{+8.3}_{-12.1}$ & 2, 8 \\
PG 1229+204& 0.063 & 7.84& 45.11& $ -1.02 \pm 0.02 $ & $37.8^{+27.6}_{-15.3}$ & 2, 7 \\
PG 1244+026& 0.048 & 6.50& 44.30& $ -0.49 \pm 0.05 $ & \nodata & 2 \\
PG 1259+593& 0.472 & 8.90& 46.76& $ -0.04 \pm 0.11 $ & \nodata & 2 \\
PG 1302$-$102& 0.286 & 8.86& 46.45& $ -0.05 \pm 0.01 $ & \nodata & 2 \\
PG 1307+085& 0.155 & 8.62& 45.80& $ -0.25 \pm 0.03 $ & $105.6^{+36.0}_{-46.6}$ & 2, 7 \\
PG 1310$-$108& 0.035 & 7.86& 44.13& $ -0.59 \pm 0.06 $ & \nodata & 2 \\
PG 1322+659& 0.168 & 8.26& 45.52& $ -0.22 \pm 0.02 $ & \nodata & 2 \\
PG 1341+258& 0.087 & 8.02& 44.71& $ -0.19 \pm 0.08 $ & \nodata & 2 \\
PG 1351+695& 0.030 & 7.52& 43.63& $ -0.74 \pm 0.06 $ & \nodata & 2 \\
PG 1352+183& 0.158 & 8.40& 45.60& $ -0.20 \pm 0.04 $ & \nodata & 2 \\
PG 1402+261& 0.164 & 7.92& 45.95& $ -0.66 \pm 0.03 $ & \nodata & 2 \\
PG 1404+226& 0.098 & 6.87& 44.86& $ -0.14 \pm 0.07 $ & \nodata & 2 \\
PG 1415+451& 0.114 & 7.99& 45.29& $ -0.10 \pm 0.03 $ & \nodata & 2 \\
PG 1416$-$129& 0.129 & 9.02& 44.93& $ -1.00 \pm 0.06 $ & \nodata & 2 \\
PG 1425+267& 0.366 & 9.71& 46.15& $ -0.76 \pm 0.02 $ & \nodata & 2 \\
PG 1426+015& 0.086 & 9.09& 45.63& $ -0.44 \pm 0.06 $ & $95.0^{+29.9}_{-37.1}$ & 2, 7 \\
PG 1427+480& 0.221 & 8.07& 45.75& $ -0.49 \pm 0.03 $ & \nodata & 2 \\
PG 1434+590& 0.031 & 7.77& 44.93& $ -0.31 \pm 0.02 $ & \nodata & 2 \\
PG 1435$-$067& 0.129 & 8.34& 45.56& $ -0.29 \pm 0.04 $ & \nodata & 2 \\
PG 1440+356& 0.077 & 7.45& 45.59& $ -0.09 \pm 0.03 $ & \nodata & 2 \\
PG 1444+407& 0.267 & 8.27& 46.24& $ 0.04 \pm 0.04 $ & \nodata & 2 \\
PG 1448+273& 0.065 & 6.95& 44.36& $ -0.12 \pm 0.06 $ & \nodata & 2 \\
PG 1501+106& 0.036 & 8.50& 44.51& $ -1.09 \pm 0.02 $ & \nodata & 2 \\
PG 1512+370& 0.371 & 9.35& 46.36& $ -0.33 \pm 0.08 $ & \nodata & 2 \\
PG 1519+226& 0.137 & 7.92& 45.16& $ -0.34 \pm 0.08 $ & \nodata & 2 \\
PG 1534+580& 0.030 & 7.37& 44.16& $ -0.66 \pm 0.02 $ & \nodata & 2 \\
PG 1543+489& 0.400 & 7.98& 46.26& $ 0.04 \pm 0.02 $ & \nodata & 2 \\
PG 1545+210& 0.266 & 9.29& 45.98& $ -0.44 \pm 0.03 $ & \nodata & 2 \\
PG 1552+085& 0.119 & 7.52& 44.81& $ -0.38 \pm 0.09 $ & \nodata & 2 \\
PG 1612+261& 0.131 & 8.04& 45.07& $ -0.79 \pm 0.06 $ & \nodata & 2 \\
PG 1613+658& 0.129 & 8.43& 45.94& $ -0.44 \pm 0.01 $ & $40.1^{+15.0}_{-15.2}$ & 2, 7 \\
PG 1617+175& 0.112 & 8.75& 45.24& $ -0.52 \pm 0.04 $ & $71.5^{+29.6}_{-33.7}$ & 2 \\
PG 1626+554& 0.133 & 8.48& 45.72& $ -0.16 \pm 0.03 $ & \nodata & 2 \\
PG 2112+059& 0.466 & 8.98& 46.25& $ 0.17 \pm 0.08 $ & \nodata & 2 \\
PG 2130+099& 0.063 & 8.64& 44.92& $ -0.54 \pm 0.03 $ & \nodata & 2 \\
PG 2214+139& 0.067 & 8.53& 44.39& $ -0.96 \pm 0.12 $ & \nodata & 2 \\
PG 2233+134& 0.325 & 8.02& 46.16& $ 0.11 \pm 0.04 $ & \nodata & 2 \\
PG 2251+113& 0.323 & 8.97& 45.83& $ -0.15 \pm 0.05 $ & \nodata & 2 \\
PG 2304+042& 0.042 & 8.54& 43.72& $ -0.85 \pm 0.06 $ & \nodata & 2 \\
PG 2308+098& 0.432 & 9.57& 46.33& $ -0.20 \pm 0.09 $ & \nodata & 2 \\
2QZ J001221.1$-$283630 & 2.327 & 8.71 & 47.11  & $-$0.08& \nodata & 3 \\
2QZ J002830.4$-$281706 & 2.406 & 9.71 & 47.43  & $-$0.33& \nodata & 3 \\
UM 667 & 3.122 & 9.15 & 47.13  & $-$0.09& \nodata & 3 \\
LBQS 0109+0213 & 2.343 & 10.00 & 47.65  & $-$0.09& \nodata & 3 \\
$[\rm HB89]$ 0123+257 & 2.358 & 9.09 & 47.43  & $-$0.11& \nodata & 3 \\
2QZ J023805.8$-$274337 & 2.452 & 9.40 & 47.42  & $-$0.05& \nodata & 3 \\
SDSS J024933.42$-$083454.4 & 2.491 & 9.66 & 47.23  & 0.03& \nodata & 3 \\
$[\rm HB89]$ 0329$-$385 & 2.423 & 10.11 & 47.56  & $-$0.07& \nodata & 3 \\
$[\rm HB89]$ 0504+030 & 2.463 & 8.80 & 47.17  & $-$0.32& \nodata & 3 \\
SDSS J100428.43+001825.6 & 3.046 & 9.33 & 47.29  & 0.18& \nodata & 3 \\
TON 618 & 2.219 & 10.82 & 48.16  & 0.20& \nodata & 3 \\
$[\rm HB89]$ 1246$-$057 & 2.236 & 10.22 & 48.01  & 0.26& \nodata & 3 \\
$[\rm HB89]$ 1318$-$113 & 2.308 & 9.75 & 47.74  & $-$0.25& \nodata & 3 \\
$[\rm HB89]$ 1346$-$036 & 2.344 & 9.94 & 47.73  & $-$0.11& \nodata & 3 \\
SDSS J135445.66+002050.2 & 2.511 & 9.13 & 47.34  & 0.47& \nodata & 3 \\
UM 629 & 2.462 & 9.16 & 47.41  & $-$0.10& \nodata & 3 \\
UM 632 & 2.518 & 9.43 & 47.39  & $-$0.55& \nodata & 3 \\
UM 642 & 2.372 & 9.35 & 47.14  & $-$0.29& \nodata & 3 \\
UM 645 & 2.269 & 9.37 & 47.16  & $-$0.54& \nodata & 3 \\
SBS 1425+606 & 3.160 & 9.82 & 48.23  & $-$0.33& \nodata & 3 \\
SDSS J170102.18+612301.0 & 2.293 & 9.72 & 47.19  & 0.08& \nodata & 3 \\
SDSS J173352.22+540030.5 & 3.425 & 9.57 & 47.85  & 0.25& \nodata & 3 \\
$[\rm HB89]$ 2126$-$158 & 3.268 & 9.72 & 48.10  & $-$0.11& \nodata & 3 \\
$[\rm HB89]$ 2132+014 & 3.194 & 8.65 & 46.62  & $-$0.10& \nodata & 3 \\
2QZ J221814.4$-$300306 & 2.384 & 9.27 & 47.39  & $-$0.02& \nodata & 3 \\
2QZ J222006.7$-$280324 & 2.406 & 10.16 & 48.07  & 0.22& \nodata & 3 \\
$[\rm HB89]$ 2254+024d & 2.081 & 9.09 & 47.30  & $-$0.23& \nodata & 3 \\
2QZ J231456.8$-$280102 & 2.392 & 9.26 & 47.16  & 0.08& \nodata & 3 \\
2QZ J234510.3$-$293155 & 2.360 & 9.37 & 47.17  & 0.20& \nodata & 3 \\
SDSS J082854.44+571637.2 & 3.383 & 9.08 & 46.62 & $ -0.21 \pm 0.10 $ & \nodata & 4 \\
SDSS J124652.80+545140.6 & 3.360 & 9.15 & 46.65 & $ -0.04 \pm 0.15 $ & \nodata & 4 \\
SDSS J120308.69+552245.8 & 3.355 & 8.83 & 46.62 & $ 0.15 \pm 0.10 $ & \nodata & 4 \\
SDSS J081528.12+344737.0 & 3.200 & 9.15 & 46.69 & $ -0.08 \pm 0.06 $ & \nodata & 4 \\
SDSS J095617.14+373224.7 & 3.245 & 8.39 & 46.39 & $ -0.08 \pm 0.09 $ & \nodata & 4 \\
SDSS J133600.20+391826.2 & 3.228 & 8.91 & 46.20 & $ -0.17 \pm 0.12 $ & \nodata & 4 \\
SDSS J142903.86+062620.4 & 3.268 & 9.06 & 46.38 & $ -0.31 \pm 0.16 $ & \nodata & 4 \\
SDSS J010049.76+092936.1 & 3.118 & 9.55 & 47.22 & $ 0.10 \pm 0.03 $ & \nodata & 4 \\
SDSS J013735.46$-$004723.4 & 3.209 & 9.08 & 46.97 & $ -0.01 \pm 0.04 $ & \nodata & 4 \\
SDSS J093514.41+343659.5 & 3.227 & 9.23 & 47.19 & $ -0.01 \pm 0.03 $ & \nodata & 4 \\
SDSS J223408.99+000001.6 & 3.026 & 9.42 & 47.82 & $ 0.12 \pm 0.01 $ & \nodata & 4 \\
SDSS J231934.77$-$104037.0 & 3.170 & 9.08 & 47.39 & $ 0.07 \pm 0.03 $ & \nodata & 4 \\
J0008$-$0626 & 5.929 & 9.19 & 46.96  & $ 0.00 \pm 0.08 $ & \nodata & 5 \\
J0028+0457 & 5.982 & 9.91 & 46.97  & $ 0.03 \pm 0.11 $ & \nodata & 5 \\
J0050+3445 & 6.251 & 9.76 & 47.10  & $ -0.11 \pm 0.08 $ & \nodata & 5 \\
J0810+5105 & 5.805 & 9.29 & 47.19  & $ 0.15 \pm 0.07 $ & \nodata & 5 \\
J0835+3217 & 5.902 & 8.91 & 46.29  & $ -0.12 \pm 0.05 $ & \nodata & 5 \\
J0836+0054 & 5.834 & 9.61 & 47.62  & $ 0.19 \pm 0.03 $ & \nodata & 5 \\
J0841+2905 & 5.954 & 9.40 & 46.99  & $ 0.01 \pm 0.08 $ & \nodata & 5 \\
J0842+1218 & 6.069 & 9.52 & 47.20  & $ -0.07 \pm 0.03 $ & \nodata & 5 \\
J1044$-$0125 & 5.780 & 9.81 & 47.31  & $ 0.02 \pm 0.03 $ & \nodata & 5 \\
J1137+3549 & 6.009 & 9.76 & 47.28  & $ -0.01 \pm 0.04 $ & \nodata & 5 \\
J1143+3808 & 5.800 & 9.73 & 47.00  & $ 0.00 \pm 0.05 $ & \nodata & 5 \\
J1148+0702 & 6.344 & 9.38 & 47.09  & $ -0.12 \pm 0.04 $ & \nodata & 5 \\
J1250+3130 & 6.138 & 9.13 & 46.99  & $ 0.00 \pm 0.03 $ & \nodata & 5 \\
J1436+5007 & 5.809 & 9.30 & 47.04  & $ 0.04 \pm 0.06 $ & \nodata & 5 \\
J1545+6028 & 5.794 & 9.20 & 46.53  & $ 0.01 \pm 0.04 $ & \nodata & 5 \\
J1609+3041 & 6.146 & 9.44 & 46.65  & $ -0.04 \pm 0.04 $ & \nodata & 5 \\
J1623+3112 & 6.254 & 9.32 & 46.98  & $ -0.04 \pm 0.04 $ & \nodata & 5 \\
J1630+4012 & 6.066 & 9.27 & 46.76  & $ -0.08 \pm 0.06 $ & \nodata & 5 \\
P060+24 & 6.170 & 9.32 & 47.06  & $ 0.04 \pm 0.07 $ & \nodata & 5 \\
J0921+0007 & 6.565 & 8.42  & 46.78 & $ 0.49 \pm 0.13 $ & \nodata & 6 \\
J0024+3913 & 6.773 & 8.43  & 46.90 & $ 0.77 \pm 0.15 $ & \nodata & 6 \\
J0910+1656 & 6.719 & 8.61  & 47.00 & $ 0.50 \pm 0.08 $ & \nodata & 6 \\
J0829+4117 & 6.773 & 8.85  & 47.10 & $ 0.49 \pm 0.09 $ & \nodata & 6 \\
J2102$-$1458 & 6.652 & 8.87  & 46.78 & $ 1.82 \pm 0.44 $ & \nodata & 6 \\
J0837+4929 & 6.702 & 8.91  & 47.16 & $ 1.49 \pm 0.08 $ & \nodata & 6 \\
P333+26 & 6.027 & 8.93  & 46.90 & $ 0.80 \pm 0.47 $ & \nodata & 6 \\
PSOJ323+12 & 6.586 & 8.91  & 47.26 & $ 0.73 \pm 0.05 $ & \nodata & 6 \\
PSOJ359$-$06 & 6.172 & 9.00  & 47.30 & $ 0.96 \pm 0.13 $ & \nodata & 6 \\
PSOJ308$-$27 & 5.799 & 9.08  & 47.32 & $ 0.98 \pm 0.06 $ & \nodata & 6 \\
ATLASJ029$-$36 & 6.020 & 9.28  & 47.25 & $ 1.07 \pm 0.11 $ & \nodata & 6 \\
PSOJ029$-$29 & 5.976 & 9.34  & 47.49 & $ 1.14 \pm 0.13 $ & \nodata & 6 \\
PSOJ158$-$14 & 6.068 & 9.36  & 47.67 & $ 0.81 \pm 0.07 $ & \nodata & 6 \\
PSOJ242$-$12 & 5.830 & 9.79  & 47.35 & $ 0.87 \pm 0.11 $ & \nodata & 6 \\
VDESJ0224$-$4711 & 6.528 & 9.20  & 47.56 & $ 0.79 \pm 0.04 $ & \nodata & 6 \\
PSOJ060+24 & 6.170 & 9.22  & 47.36 & $ 0.76 \pm 0.05 $ & \nodata & 6 \\
\enddata
\tablecomments{References: 1. \citet{du2014}. 2. \citet{shin2013}. 3. \citet{shemmer2004}. 4. \citet{shin2019}. 5. \citet{wang2022}. 6. \citet{lai2022}. 7. \citet{aldama2019}. 8. \citet{du2019}.}
\end{deluxetable*}

\end{document}